\documentclass[twocolumn,prb,showpacs,preprintnumbers,amsmath,amssymb]{revtex4-1}

\usepackage{graphicx}

\begin{document}


\title{Orbital-dependent electron correlation effect on the two- and three-dimensional Fermi surfaces in KFe$_2$As$_2$ revealed by angle-resolved photoemission
spectroscopy}

\author{T. Yoshida$^{1,2}$, S. Ideta$^1$, I. Nishi$^1$, A. Fujimori$^{1,2}$, M.
Yi$^3$, R. G. Moore$^3$, S. K. Mo$^3$, D.-H. Lu$^3$, Z.-X.
Shen$^3$, Z. Hussain$^4$, K. Kihou$^{2,5}$, P. M. Shirage$^{2,5}$,
H. Kito$^{2,5}$, C. H. Lee$^{2,5}$, A. Iyo$^{2,5}$, H.
Eisaki$^{2,5}$, H. Harima$^{2,6}$}

\affiliation{$^1$Department of Physics, University of Tokyo,
Bunkyo-ku, Tokyo 113-0033, Japan}

\affiliation{$^2$JST, Transformative Research-Project on Iron
Pnictides (TRIP), Chiyoda, Tokyo 102-0075, Japan}

\affiliation{$^3$Department of Applied Physics and Stanford
Synchrotron Radiation Laboratory, Stanford University, Stanford,
CA94305, USA}

\affiliation{$^4$Advanced Light Source, Lawrence Berkeley National
Lab, Berkeley, California 94720, USA}

\affiliation{$^5$National Institute of Advanced Industrial Science
and Technology (AIST), Tsukuba 305-8562, Japan}

\affiliation{$^6$Department of Physics, Kobe University, Kobe,
Hyogo 657-8561, Japan}

\date{\today}

\begin{abstract}
We have performed an angle-resolved photoemission study of the
hole-overdoped iron pnictide superconductor KFe$_2$As$_2$, which
shows a low $T_c$ of $\sim$4 K. Most of the observed Fermi
surfaces show nearly two-dimensional shapes, while a band near the
Fermi level shows a strong dispersion along the $k_z$ direction
and forms a small three-dimensional hole pocket centered at the
$Z$ point, as predicted by band-structure calculation. However,
hole Fermi surfaces of $yz$ and $zx$ orbital character centered at
the $\Gamma$ point of the two-dimensional Brillouin zone are
smaller than those predicted by the calculation while the other
hole Fermi surfaces of $xy$ orbital character is much larger.
Clover-shaped hole Fermi surfaces around the corner of the 2D BZ
are also larger than those predicted by the calculation. These
observations are consistent with the de Haas-van Alphen
measurement and indicate orbital-dependent electron correlation
effects. The effective masses of the energy bands show moderate to
strong enhancement, partly due to electron correlation and partly
due to energy shifts from the calculated band structure.
\end{abstract}


\maketitle

\section{INTRODUCTION}
In contrast to the $d$-wave superconducting gaps in the high-$T_c$
cuprate superconductors, experimental results on most of the
iron-pnictide superconductors have indicated that superconducting
gaps are nodeless and on the entire Fermi surfaces (FSs)
\cite{Ding, Terashima}. However, some of the iron pnictide
superconductors show signatures of the nodes in the
superconducting gaps. For example, thermal conductivity
measurements of isovalent substituted system
BaFe$_2$(As$_{1-x}$P$_x$)$_2$ \cite{Hashimoto} and the electron
doped systems Ba(Fe$_{1-x}$Co$_x$)$_2$As$_2$ and
Ba(Fe$_{1-x}$Ni$_x$)$_2$As$_2$ \cite{Reid} in the superconducting
state have shown signature of line nodes. According to the
theories of spin fluctuation-mediated superconductivity, line
nodes may appear when the pnictogen height becomes small
\cite{Kuroki, Ikeda}, the hole FS of $d_{xy}$ character around the
zone center disappears and nesting between hole and electron FSs
becomes weakened. (Here, $x$ and $y$ are referred to the direction
of the nearest neighbor Fe atoms.) The hole FSs of these systems
exhibit strong three-dimensionality \cite{YoshidaAsP, Malaeb,
Vilmercati}, resulting in poor nesting between the hole and
electron FSs.

In the K-doped BaFe$_2$As$_2$ (Ba122) system, a full
superconducting gaps opens and a high $T_c$ of $\sim$ 37 K is
achieved in the optimally doped region \cite{Ding}, while the
existence of line nodes in the superconducting gap has been
suggested for the end member compound KFe$_2$As$_2$, which shows a
low $T_c$ of $\sim$ 4 K \cite{Terashima} from penetration depth
\cite{Hashimoto_K122}, thermal conductivity \cite{Dong}, and
nuclear quadrupole resonance (NQR) measurements \cite{Fukazawa}.
The electronic specific heat coefficient $\gamma$ of KFe$_2$As$_2$
is as large as $\sim$70 mJ/K$^2$mol \cite{Fukazawa}, indicating
strong electron mass renormalization due to electron correlation.
A previous angle-resolved photoemission study (ARPES) of
KFe$_2$As$_2$ has revealed that there are hole FSs ($\alpha$ and
$\beta$) around the zone center while the electron pockets around
the zone corner disappear and change to small hole pockets
($\epsilon$) surrounding the zone corner in a clover shape
\cite{Sato}. Band-structure calculation has predicted the third
hole FS around the zone center, namely, the $\zeta$ hole FS which
has a similar size to the $\alpha$ FS, but it has not been
resolved in the previous ARPES study. Recently, a de Haas-van
Alphen (dHvA) study has revealed that the $\zeta$ FS exists around
the zone center \cite{dHvA}. Furthermore, the dHvA study has
indicated the shrinkage of the $\alpha$ and $\zeta$ FSs and the
enhancement of the electron masses compared to those predicted by
the band-structure calculation. In the present study, in order to
reveal the shapes of the FSs in three-dimensional momentum space,
we have performed an ARPES study of KFe$_2$As$_2$ using
high-quality single crystals and various photon energies. We have
determined the orbital character of the FSs by polarization
dependent measurements and have revealed strongly orbital
dependent correlation effects.

\begin{figure}[htbp]
 \includegraphics[width=8.5cm]{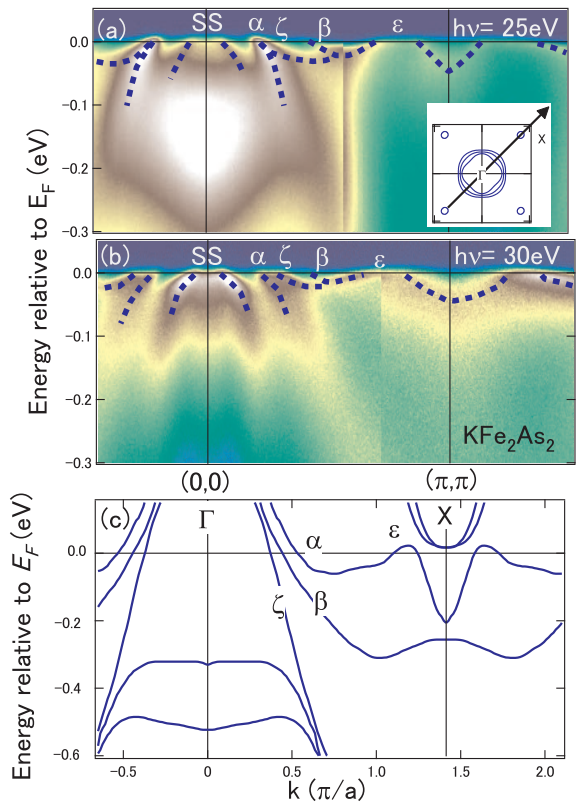}
 \caption{Band dispersions of KFe$_2$As$_2$ in the zone diagonal direction.
(a),(b) ARPES spectra taken at h$\nu$= 25eV and 30eV,
respectively, corresponding to $k_z$=6.5 and 7.0 (2$\pi$/$c$). SS
denotes surface states. (c) Band dispersions predicted by
band-structure calculation.}\label{Ek}
\end{figure}

\section{EXPERIMENT AND BAND-STRUCTURE CALCULATION}
ARPES measurements were performed at beamline 5-4 of Stanford
Synchrotron Radiation Lightsource (SSRL) and at BL10.0.1 of
Advanced Light Source (ALS). Scienta SES-R4000 electron analyzers
and linearly polarized light were used at both beamlines. The
typical energy resolutions were 10 meV at SSRL and 20 meV at ALS,
respectively. Single crystals of KFe$_2$As$_2$ were grown from a K
flux. Samples were cleaved \textit{in situ} and measured at a
temperature of 15 K in a pressure better than $5 \times 10^{-11}$
Torr. We have performed the measurements at photon energies from
$h\nu$=14 to 40 eV. The in-plane ($k_X$, $k_Y$) and out-of-plane
($k_Z$) momentum are expressed in units of $\pi/a$ and $2\pi/c$,
respectively, where $a= 3.864$ \textrm{\AA} and $c=13.87$
\textrm{\AA}. Here, the $X$ and $Y$ axes point towards the Fe-As
bond direction, while the $x$ and $y$ axes are rotated by 45
degrees from the $X$-$Y$ coordination. The electronic band
structure of KFe$_2$As$_2$ was calculated within the local density
approximation (LDA) by using the full potential LAPW (FLAPW)
method. We used the program codes TSPACE \cite{Yanase} and
KANSAI-06. The experimental crystal structure \cite{Rozsa}
including the atomic position $z_\mathrm{As}$ of As (pnictogen
height) was used for the calculation.

\begin{figure}
\includegraphics[width=9cm]{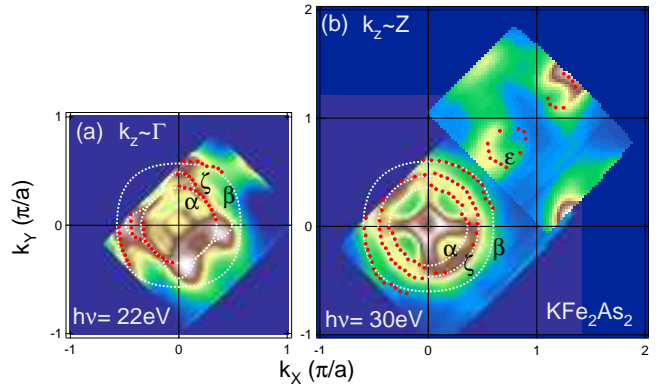}
\caption{\label{FS2D}(Color online) FSs of KFe$_2$As$_2$ observed
by ARPES. ARPES intensity at $E_F$ mapped in the $k_X$-$k_Y$ plane
taken at two different photon energies. Red dots indicate $k_F$
positions determined by the peak positions of momentum
distribution curves (MDC's) and white dotted lines indicate FSs
deduced from the $k_F$ positions.}
\end{figure}

\section{RESULTS AND DISCUSSION}
Band dispersions for a cut along the diagonal of the
two-dimensional Brillouin zone (BZ) taken with $h\nu$=25 eV and
30eV are shown in Figs. \ref{Ek}(a) and \ref{Ek}(b). All the
energy bands predicted by the calculation [Fig. \ref{Ek}(c)] are
observed. Particularly, we have clearly observed the $\zeta$ band
near the $\alpha$ band around $E_F$ consistent with the dHvA
result \cite{dHvA}. In addition to these bands, we find that
another hole-like band crossing $E_F$ exists near the zone center.
Since this is not predicted by the bulk band-structure
calculation, we attributed this band to surface states. While the
$\zeta$ band is nearly degenerated with the $\alpha$ band for
h$\nu$ =25 eV, these bands are separated for h$\nu$ =30 eV,
indicating three-dimensionality of the band dispersions.

FS mapping in the $k_X$-$k_Y$ plane is shown in Figs. \ref{FS2D}
(a) and \ref{FS2D}(b). By assuming the inner potential $V_0$=13.0
eV, panels (a) and (b) approximately represent the $k_X$-$k_Y$
planes including the $\Gamma$ and the $Z$ point, respectively. The
overall FS shapes nearly agree with those observed in the previous
study \cite{Sato}. Small hole FSs also appear around the BZ corner
due to heavy hole doping. However, several new observations should
be remarked. One is that all the three hole FSs around the center
of the 2D BZ have been clearly resolved. The middle hole FS is
ascribed to the $\zeta$ FS reported in the dHvA study \cite{dHvA}.
Another point is that a small FS around the zone center has been
observed. Since this FS is not predicted by the band-structure
calculation and has nearly two-dimensional dispersions as
indicated below, it can be ascribed to surface states. The
dispersions of the surface states form ridge-like structures
extending to the $k_X$ and $k_Y$ directions, resulting in the
peculiar cross-like intensity distribution at $E_F$.

\begin{figure}
\includegraphics[width=8.5cm]{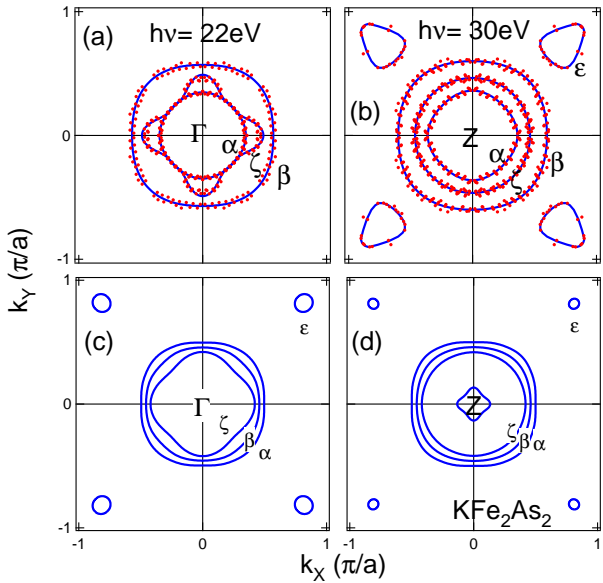}
\caption{\label{FS}(Color online) Comparison of FSs obtained by
ARPES and those predicted by band-structure calculation. (a),(b)
FSs determined by ARPES. $k_F$ positions in Fig. \ref{FS2D} have
been symmetrized in the first BZ. (c),(d) FSs given by the
band-structure calculation. A small FS around the $Z$ point comes
from a three-dimensional $d_{z^2}$ band shown in Fig. \ref{FS3D}
(c).}
\end{figure}

In Fig.\ref{FS}, we compare the FSs obtained by ARPES with the
band-structure calculation. As seen in panels (a) and (b), the
sizes of the observed $\alpha$ and $\beta$ FSs do not show
appreciable change with $k_z$. On the other hand, the shape of the
$\zeta$ FS significantly changes between $k_z\sim\Gamma$ and $Z$.
While the $\zeta$ FS has a diamond-like cross-section for
$k_z\sim\Gamma$ and is nearly degenerate with the $\alpha$ FS in
the zone diagonal direction, it has a circular cross-section for
$k_z\sim Z$ point.

\begin{figure}
\includegraphics[width=8.5cm]{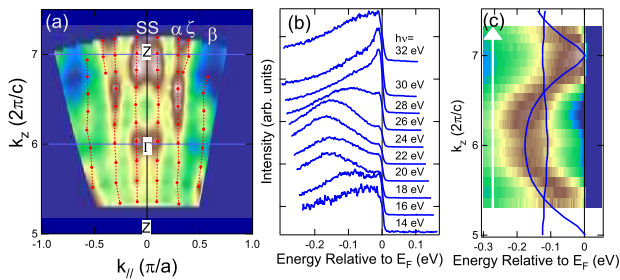}
\caption{\label{FS3D}(Color online) $k_z$ dependence of the
electronic structure of KFe$_2$As$_2$. (a) Spectral weight mapping
at $E_F$ in the $k_\parallel$-$k_z$ plane, where $k_\parallel$ is
in the $\Gamma$-$X$ direction. The peak positions of momentum
distribution curves are shown by red circles. SS denotes surface
states. (b) Normal emission spectra corresponding to the arrow in
panel (a). (c) ARPES intensity plot along the $\Gamma$-$Z$ line.
Three-dimensional $d_{3z^2-r^2}$ band predicted by the
band-structure calculation with renormalized mass $m^*/m_b$ =3 are
also plotted. }
\end{figure}

Since the Ba122 system in general has three-dimensional FSs as
predicted by band-structure calculations \cite{Singh} and
confirmed by ARPES \cite{Malaeb,Vilmercati}, we have investigated
the electronic band dispersions in the $k_z$ direction by changing
the excitation photon energy. Figure \ref{FS3D}(a) shows spectral
weight mapping at $E_F$ in the $k_\parallel$-$k_z$ plane, where
the direction $k_\parallel$ in the $k_X$-$k_Y$ plane is the
$\Gamma$-$X$ direction. $k_z$ has been determined by assuming the
inner potential of $V_0$= 13.0 eV, so that the dispersion of the
$d_{3z^2-r^2}$ band along $k_z$ has the correct periodicity and
phase as mentioned below. From Fig. \ref{FS3D}(a), one can see
that the $\alpha$ and $\beta$ FSs are nearly two-dimensional,
while the $\zeta$ FS has some three dimensionality and becomes
large near the $Z$ point compared to the $\Gamma$ point. This
behavior has also been observed in the ARPES spectrum of K-doped
Ba122 \cite{Zhang3D}. In normal emission spectra [Fig.\ref{FS3D}
(b)] the $d_{3z^2-r^2}$ band clearly shows a parabolic dispersion
along the $k_z$ direction and slightly crosses $E_F$ in the
vicinity of the $Z$ point forming a small three-dimensional hole
pocket around the $Z$ point. With a mass enhancement factor
$m^*/m_b\sim$ 3, this band is also well explained by the
band-structure calculation.

\begin{figure}
\includegraphics[width=8.5cm]{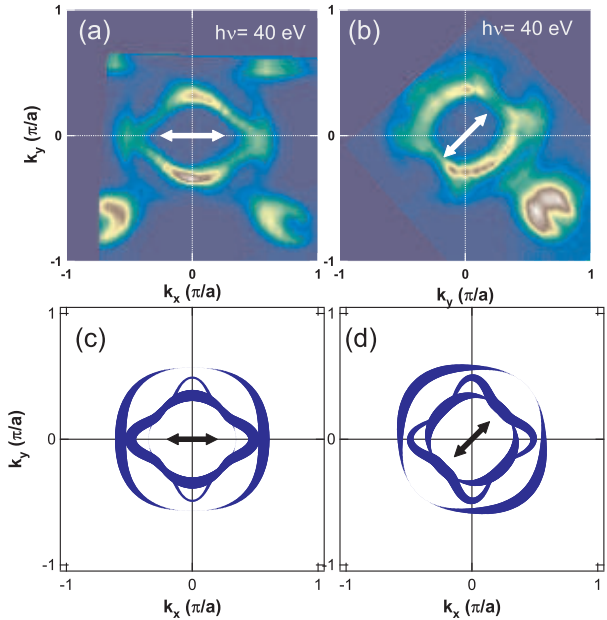}
\caption{\label{ME}(Color online) Polarization dependence of the
FS mapping for KFe$_2$As$_2$. (a),(b) Measured ARPES intensity at
$E_F$ in the $k_X$-$k_Y$ plane taken at $h\nu$=40 eV ($k_z\sim
\Gamma$ ). Electric vectors are shown by arrows. (c),(d)
Simulation of the ARPES intensity distribution corresponding to
panels (a) and (b). Shapes of FSs have been taken from Fig.
\ref{FS2D}. By assuming certain orbital character for each FS,
intensity distribution has been simulated and is shown by
thickness of the curves. (For details, see the text.) }
\end{figure}

In order to determine the orbital character of the FSs, we have
investigated the polarization dependence of the ARPES intensity as
shown in Fig. \ref{ME}. FS mapping shown in panels (a) and (b)
indicates clear polarization dependence in the intensity
distribution for each FS. We have simulated the intensity
distribution by using the following assumptions. Based on the
result of  the band-structure calculation, we assume that three
orbitals $xy$, $yz$ and $zx$ constitute the FSs. We refer to the
three band as $xy$, $yz$ and $zx$ band according to the orbital
character of the band with momentum in the zone diagonal $k_x$
($\parallel k_X$+$k_Y$) direction. Using the angle $\theta$ around
the $\Gamma$ point, the orbital character of the $xy$, $zx$ and
$yz$ band can be approximately expressed by $|xy>$,
$\cos\theta|zx> + \sin\theta|yz>$ and $-\sin\theta|zx> +
\cos\theta|yz>$, respectively. By assuming the dipole
approximation of the transition matrix element
$|<i|\mathbf{\varepsilon}\cdot\mathbf{r}|f>|^2$, where $|i>$,
$|f>$, and $\mathbf{\varepsilon}$ are the initial state, the final
state, and the polarization vector, respectively, one can predict
the intensity distribution. For example, when $|i>=|xy>$,
$\mathbf{\varepsilon}//x$ and $|f>$ is a wave function of a free
electron, the transition matrix element
$|<i|\mathbf{\varepsilon}\cdot\mathbf{r}|f>|^2$ is proportional to
$k_y^2$ in the lowest order in $\mathbf{k}$.

Figures \ref{ME}(c) and \ref{ME}(d) are the results of the
intensity simulations. Here, we assign the inner, middle, and
outer FSs to the $yz$, $xz$, and $xy$-band, respectively, so that
we can reproduce the experimental intensity distribution. This
assignment of the orbital character is difference from the
band-structure calculation where the inner, middle, and outer FSs
have $xz$, $xy$ and $yz$ orbital character, respectively. However,
present ARPES result is consistent with the previous ARPES result
of Co-Ba122 \cite{Zhang} and the theoretical prediction of
LDA+DMFT \cite{Yin}, which indicate the energy inversion of the
$xy$ and $yz/xz$ bands due to orbital-dependent correlation
effect. That is, the $xy$ band in most strongly affected by
electron correlation and is shifted upward relative to the other
bands.

Another discrepancy from the band-structure calculation is the
inversion of the $yz$ and $xz$ bands in the $k_X$ ($k_Y$)
direction. In the result of the band-structure calculation, the
inner FS has $xz$ character (in the $k_X$ direction) with
rounded-square shape around the $\Gamma$ point and becomes
circular around the $Z$ point because of hybridization with the
$z^2$ orbital. In the present ARPES result, such a character has
been observed in the middle $\zeta$ FS. According to the
angular-dependent magnetoresistance oscillations, such a
rounded-square FS is also bigger than a circular hole FS
\cite{Kimata}. The observed inversion of the $xz$, $yz$ bands is
consistent with the ARPES result of Co-Ba122 \cite{Zhang} and the
LDA+DMFT calculation for KFe$_2$As$_2$ \cite{Yin}.

In the dHvA study, the size of the hole FSs are found to be
smaller than those predicted by band-structure calculations
\cite{Terashima}. We have determined the cross-sectional area of
the FSs as listed in Table \ref{masstable} together with those of
the dHvA measurements and the band-structure calculation. The
cross-sectional areas for the $\alpha$ and $\zeta$ FSs observed by
ARPES are close to those obtained by the dHvA result and are
smaller than the band-structure calculation. On the other hand,
the area of the $\beta$ and $\epsilon$ FSs determined by ARPES are
much larger than the calculation results. The total hole count
from the observed FSs yields the hole FS volume of 61 \% of the
BZ, indicating a deviation from the value of 50\% expected from
the chemical composition, because most of the FSs observed by
ARPES are nearly 10-20 \% larger than those observed by dHvA. The
deviation of the FS volume implies that there are excess hole
doping of 0.11 per Fe atom at the sample surface. Nevertheless the
surface effect is not so serious as those in 1111 system where
excess 0.5-0.6 holes per Fe are doped, and one can still discuss
mass renormalization from the present result.

\begin{table}[htbp]
\caption{Cross-sectional areas and effective masses of FSs of
KFe$_2$As$_2$ determined by ARPES and dHvA experiment. The areas
are expressed as a percentage of the area of the 2D BZ. $m_e$ and
$m_b$ are the free-electron and band masses, respectively.}
  \label{masstable}
\begin{tabular}{lc|lcc|ccc}
        \hline\hline
    FS   & $k_z$ &   &area&   &  & $m^*/m_e$ ($m^*/m_b$)  &    \\
         &          &ARPES&dHvA & LDA &ARPES&dHvA& LDA \\
\hline
$\alpha$ & $\Gamma$ & 9.1 & 8.2   & 20.8 & 5.1 (2.0) & 6.0 (2.3) & 2.6  \\
         & $Z$      & 9.8 & 8.6   & 21.6 & 6.6 (2.3) & 6.5 (2.2) & 2.9  \\
$\zeta$  & $\Gamma$ & 12.2 & 10.3 & 12.2 & 11.0 (7.9) & 8.5 (6.1) & 1.4  \\
         & $Z$      & 17.0 & 15.7 & 13.8 & 9.6 (4.0) & 18 (7.5) & 2.4 \\
$\beta$  & $\Gamma$ & 27.3 &      & 16.7 & 16.3 (6.3) &     & 2.6 \\
         & $Z$      & 30.0 &      & 17.4 & 17.9 (6.9) &     & 2.6 \\
$\epsilon$ & $\Gamma$& 2.1 & 0.86  & 0.11 & 5.6 (18.7) & 6.0 (20) & 0.3 \\
           & $Z$     &       & 1.29  & 0.36 &  & 7.2 (24) &      0.3\\
        \hline\hline
    \end{tabular}
\end{table}

The effective masses determined by ARPES are compared with those
derived from the dHvA measurements and the band-structure
calculation in Table \ref{masstable}. For the $\alpha$ and $\zeta$
FSs, the effective mass ratio $m^*/m_e$, where $m_e$ is the free
electron mass, determined by ARPES is in good agreement with those
obtained by dHvA, except for the $k_z=Z$ data for the $\zeta$ FS.
These values give mass enhancement factor, $m^*/m_b$ of $\sim$3-4,
where $m_b$ is a band mass. As for the $\beta$ FS, $m^*/m_e$,
$\sim$16-18, corresponding to $m^*/m_b\sim$6 is much larger than
other FSs. This strong mass enhancement for the $\beta$ band is
partly due to the fact that the band bottom is closer to the Fermi
level than that of the calculation. Another origin of the
enhancement is electron correlation due to the $xy$ orbital
character. According to the LDA+DMFT calculation \cite{Yin}, a
larger mass renormalization is expected in the $xy$ band than
those of the $yz/xz$ bands. Thus, the observed mass enhancement
factors indicate moderate to strong electron correlation. From the
effective masses $m^*$ listed in Table \ref{masstable}, the
electronic specific heat coefficient $\gamma$ is calculated to be
$\gamma\sim$84 mJ/mol K$^2$, which is close to $\gamma$= 70 mJ/mol
K$^2$ estimated from specific heat measurements \cite{Fukazawa}.
Since in 2D material $\gamma$ is proportional to the sum of the
$m^*$'s of all the FSs, the $\beta$ band has a large contribution
to the enhancement of $\gamma$.

The penetration depth \cite{Hashimoto_K122} and thermal
conductivity \cite{Dong} measurements of KFe$_2$As$_2$ suggest
that line nodes exist in the superconducting gap. Because the
small hole FSs around the zone corner are too small to account for
the linear temperature dependence of the superfluid density, the
node should be on the zone-centered hole FSs.  According to
spin-fluctuation mediated pairing mechanism, $s$-wave and $d$-wave
pairings are found to be in close competition in KFe$_2$As$_2$
\cite{Suzuki_K122} and they might show crossover with doping
\cite{Thomale}. For the s-wave pairing, there are two
possibilities: horizontal nodes exist on the $xz$/$yz$/$z^2$ FS or
the gap is small the entire $xy$ FS \cite{Suzuki_K122}. In the
present results, the $\beta$ hole FS which has $xy$ orbital
character is much larger than that predicted by LDA. This implies
that, if $s\pm$ paring is realized, the $\beta$ FS lays close to
the nodal line of the $s\pm$ wave order parameter, resulting in a
small superconducting gap or nodes.

Alternatively, horizontal nodes may be realized in the strongly
warped region of the FS around the $Z$ point, where a significant
amount of $3z^2$-$r^2$ character is hybridized \cite{Graser,
suzuki, Suzuki_K122}. The small-angle neutron scattering
measurement on KFe$_2$As$_2$ has suggested the existence of a
horizontal node \cite{Kawano-Furukawa}. If this is the case, the
three-dimensionality of the $\zeta$ FS revealed by the present
work may favor a horizontal node. In order to make a conclusive
remark on the line nodes, ARPES measurement of the superconducting
gap with higher energy resolution is necessary in future studies.

\section{CONCLUSION}
We have studied FSs of KFe$_2$As$_2$ in three-dimensional momentum
space by ARPES. All FSs except for the surface states are
qualitatively consistent with the band-structure calculation.
Particularly, some three-dimensionality in the $\zeta$ FS has been
identified. The sizes of the FSs nearly accord with the dHvA
observation: the $\alpha$ and $\zeta$ FSs are smaller than those
in the band-structure calculation while the $\beta$ and $\epsilon$
FSs are larger, which may be attributed to orbital-dependent
electron-electron correlation effects.


\section*{ACKNOWLEDGMENT}
We are grateful to K. Kuroki, R. Arita, H. Fukazawa, T. Terashima
and M. Kimata for enlightening discussions. Thanks are also due to
K. Haule for showing us the result of LDA-DMFT calculations prior
to publication. This work was supported by the Japan-China-Korea
A3 Foresight Program and a Grant-in-Aid for Young Scientist (B)
(22740221) from the Japan Society for the Promotion of Science.
SSRL is operated by the US DOE Office of Basic Energy Science
Divisions of Chemical Sciences and Material Sciences.

\bibliography{KFe2As2}

\end{document}